\documentstyle[12pt]{article}
\def\beq{\begin{equation}}
\def\eeq{\end{equation}}
\def\bea{\begin{eqnarray}}
\def\eea{\end{eqnarray}}
\def\bq{\begin{quote}}
\def\eq{\end{quote}}
 
\begin{document}
\pagestyle{empty}
\begin{flushright}
{ROME prep.1140/96 \\
 hep-th/9603058}
\end{flushright}
\vspace*{5mm}
\begin{center}
{\bf  LARGE-$N$ LIMIT AND CONTACT TERMS IN UNBROKEN $YM_4$ }
\\  
\vspace*{1cm} 
{\bf M. Bochicchio} \\
\vspace*{0.5cm}
INFN Sezione di Roma \\
Dipartimento di Fisica, Universita' di Roma `La Sapienza' \\
Piazzale Aldo Moro 2 , 00185 Roma  \\ 
\vspace*{2cm}  
{\bf ABSTRACT  } \\
\end{center}
\vspace*{5mm}
\noindent
I characterize the structure of the master field for $F^{0}_{z \bar z}$ in
$SU(\infty)$-$YM_4$
on a product of two Riemann surfaces $Z \times W$ in the gauge
$F^{ch}_{z \bar z}=0$ as the sum of a `bulk' constant term and of
delta-like `contact' terms.\\
The contact terms may occur because the localization of the functional integral
at $N=\infty$ on
a master orbit of a constant connection under the action of singular gauge
transformations is still compatible with the large-$N$ factorization and 
translational invariance.\\ 
In addition I argue that if the gauge group is unbroken and there is a mass
gap, that is if the theory confines, the functional measure at $N=\infty$,
in the gauge $F^{ch}_{z \bar z}=0$, must be localized on the moduli space of
flat connections with punctures on $Z \times W$.
\vspace*{1cm}
\begin{flushleft}
March 1996
\end{flushleft}
\phantom{ }
\vfill
\eject

\setcounter{page}{1}
\pagestyle{plain}

\section{Contact terms and the master orbit}
                                                                      
Many years ago, it was suggested in \cite{Witten} that because of
the factorization of correlation functions at large-$N$ and translational
invariance, the functional measure for $YM$ theories at $N=\infty$ should be
localized on the gauge orbit of some constant connection $A_{\mu}$,
the master orbit:
\bea
A_{\mu}^{g}=gA_{\mu}g^{-1}-i g^{-1}\partial_{\mu}g~.
\eea
However, in this paper I show that if the gauge tranformation $g$ is not smooth,
some gauge-invariant local 
operators do not need to be constant when computed on the master orbit,
but in addition to the constant part get the contribution of some
`contact terms'.\\
It turns out that the most general structure of the contact terms for
gauge-invariant local operators at $N=\infty$, compatible
with translational invariance and factorization, is that they are ultralocal
distributions (linear combinations of delta functions and their derivatives)
localized at submanifolds
that depend on moduli that contain some copies of the translations.
For example I show in this paper that, in the gauge $F^{ch}_{z \bar z}=0$
(the superscript $^{ch}$ means the charged part with respect to the diagonal
$U(1)^{N-1}$), in
$SU(N)$-$YM_4$ on a product of two
Riemann surfaces $Z \times W$, the structure of the master field for
the neutral part of $F_{z \bar z}$, that is the eigenvalues of $F_{z \bar z}$ 
(that determine the correlation functions of all the traces of $F_{z \bar z}$ ),
may be in general the
sum of a constant `bulk' contribution and some `contact' terms that are
delta-like distributions.\\
This situation is reminiscent of topological field theories
\cite{Witten1},\cite{Witten2},\cite{Verlinde},
in which contact terms arise because of
the topological non-triviality of the gauge orbit
\cite{Imbimbo}.\\                                                  
From now on I will restrict my argument to $SU(N)$-$YM_4$ on a product of two
Riemann surfaces $Z \times W$.\\
I will assume that Eq.(1) holds for some not-necessarily smooth gauge
transformation $g$. 
Let us suppose that the gauge $F^{ch}_{z \bar z}=0$ can be reached.
Should the gauge
transformation $g$ in Eq.(1) be smooth on the entire orbit,
$F^{0}_{z \bar z}$ would simply be a constant.
However, under a singular gauge transformation $F_{z \bar z}$ transforms as:
\bea
F^g_{z \bar z}=gF_{z \bar z}g^{-1} + F_{z \bar z}( g^{-1} \partial_{z} g,
g^{-1} \partial_{\bar z}g)
\eea
where the second term represents the field strength of a connection that is
locally a pure gauge.
In the gauge $F^{ch}_{z \bar z}=0$, that allows  residual $U(1)^{N-1}$
transformations, Eq.(2) reduces to:
\bea
F^{0 \omega}_{z \bar z}=F^0_{z \bar z}+\partial_{z}\partial_{\bar z}\omega^0-
\partial_{\bar z} \partial_{z}\omega^0 ~.
\eea
The second term is zero for smooth $\omega^0$, but it is proportional to
a singular delta function
if the gauge transformation $\omega^0$ carries a non-trivial $\pi_1$, that is
if there is a magnetic vortex in the theory:
\bea
\omega^0=h^0~ Imlog(z-z_1)~.
\eea
Only $\pi_1$ needs to be considered here since the $w$-coordinates appear
as a parameter in Eq.(3).\\
I conclude that the most general structure for the master field 
of $F^{0}_{z \bar z}$ is a constant part plus a vortex condensate:
\bea
i F^0_{z \bar z}(z,w)=H^0_{0}+ \sum_{i} h^0_i(z,w) \delta^{(2)}(z-z_i(w))
\eea
or, in a
singular $U(1)^{N-1}$ gauge in which the phase of the charged connections is
multivalued, a condensate of strings. The factor of $i$ has been introduced
into Eq.(5) to make the constant $H^0_0$ real.\\
The occurrence of these `contact' terms is obviously compatible with
translational invariance since the functional measure will contain the
integration over the moduli (positions) of the `contacts'.
In the next section I show that it is also compatible with the large-$N$
factorization of the gauge invariant correlations.

\section{Contact terms and factorization}

I start presenting a slightly different argument that does not make explicit
use of the assumption that Eq.(1) holds.                
In the gauge $F^{ch}_{z \bar z}=0$, the effective action for $F^{0}_{z \bar z}$,
$\Gamma$,
\bea
Z&=&\int \exp[-S_{YM}] \delta(F^{ch}_{z \bar z}) \delta(F^{0}_{z \bar z} - H^0)
\Delta_{FP} DA DH^0 \nonumber \\ 
&=&\int \exp[-\Gamma(H^0)]  DH^0~,
\eea
defined integrating out all the other fields but $F^{0}_{z \bar z}$,
though a priori 
unknown is of order $N^2$ for $N$ large, while the integration measure 
grows as $N$. Therefore, for the functional integral in which $\Gamma$ 
occurs, the saddle-point method applies for large $N$.
Because of translational invariance the minima or the saddles of $\Gamma$
must be either constant or non-constant configurations containing the
translations
among their moduli, that is:
\bea
 H^0(x)= H^0_0+ H^0(x;[x_i])
\eea
Because of large-$N$ clustering, for the non-constant part of $H^0(x)$ I
may assume:
\bea
H^0(x;[x_i])= \sum_i H^0(x-x_i)
\eea
where for simplicity I made the unnecessary assumption that the irreducible
constituents of the master field with respect to translations $H^0(x-x_i)$ are
all of the same `type'.
Now I compute the two-point correlation function making the `ansatz'
of Eqs.(7)-(8):
\bea
&&<H^i(x)H^j(y)>= H^i H^j + H^i \frac{n}{V} \int H^j(y-x_1) dx_1+ \nonumber\\ 
&& + H^j \frac{n}{V} \int H^i(x-x_1) dx_1+ \nonumber\\
&&+\frac{1}{V^n} \int \sum_k H^i(x-x_k) \sum_{k^{'}}
H^{j}(y-x_{k^{'}})
 \prod dx_i = \nonumber\\
&&=H^i H^j + H^i \frac{n}{V} \int H^j(y-x_1) dx_1+ \nonumber\\ 
&& + H^j \frac{n}{V} \int H^i(x-x_1) dx_1+ \nonumber\\
&&+ {\frac{1}{V}} \sum_k \int H^i(x-x_k) H^j(y-x_k) dx_k+ \nonumber\\
&&+\sum_{k \neq {k^{'}}}
\frac{1}{V} \int H^i(x-x_k) dx_k
\frac{1}{V} \int H^i(y-x_{k^{'}}) dx_{k^{'}} = \nonumber\\
&&=H^i H^j + H^i \frac{n}{V} \int H^j(y-x_1) dx_1+ \nonumber\\ 
&& + H^j \frac{n}{V} \int H^i(x-x_1) dx_1+ \nonumber\\
&&+ {\frac{n}{V}} \int H^i(x-x_1) H^j(y-x_1) dx_1 + \nonumber\\
&&+\frac{n^2-n}{V^2} [\int H^i(x-x_1) dx_1]  [\int H^j(y-x_1) dx_1] 
\eea
This correlation function should be compared with the disconnected
product:
\bea
<H^i(x)> <H^i(y)>&=&
[H^i+ \frac{n}{V} \int H^i(x-x_1) dx_1] \times \nonumber\\
&&\times [H^j+ \frac{n}{V} \int H^j(y-x_1) dx_1] 
\eea
where $n$ is the number of irreducible constituents with respect to
translations.
I also assume that the limit $n \rightarrow \infty$ is taken keeping
constant the number of constituents per unit of `volume'.
The two-point correlation function factorizes only if the 
constituents 
of the master field are either a constant or an ultralocal distribution,
otherwise the non-trivial
overlap between different irreducible constituents that appears after the last
equality in Eq.(9) would imply a non-vanishing
two-point connected function.
Quite analogous formulae hold for all the other correlation functions of
$F^{0}_{z \bar z }$.\\
Since $F^{0}_{z \bar z }$ has scaling dimension two, it is not restrictive to 
assume that $F^{0}_{z \bar z}$ is a linear combination with dimensionless
coefficients
of delta-two (anomalous dimensions are expected to appear only as $\frac{1}{N}$
corrections). All the other ultralocal distributions can be obtained as
limits of linear combinations of delta, because Dirac measures are dense in the
distributions \cite{Barry-Simon}.
Now I show explicitly that the ansatz in Eq.(5) is compatible with
factorization, provided the v.e.v.'s of $h^0_i(z,w)$ factorize and are
translational invariant (for example $h^0_i(z,w)$ are constant) :
\bea
&&i^2<F^i_{z \bar z}(z_1,w_1) F^j_{z \bar z}(z_2,w_2)>
= H_0^i H_0^j + H_0^i  \frac{<\sum_k h^j_k(z_k,w_2)>}{A}+ \nonumber\\
&& + H_0^j  \frac{<\sum_k h^i_k(z_k,w_1)>}{A} + \nonumber\\
&&+\frac{1}{A^{2n}} < \int \sum_k h^i_k(z_1,w_1) \delta^{(2)}(z_1-z_k(w_1))
\prod dz_k(w_1) \times \nonumber\\
&&\times \int \sum_{k'}  h^j_{k'}(z_2,w_2) 
\delta^{(2)}(z_2-z_{k'}(w_2))
\prod dz_{k'}(w_2)>~.
\eea
For  $w_1 \neq w_2$ 
the two-point function should be compared with the disconnected
product:
\bea
i^2<F^i_{z \bar z}(z_1,w_1) > <F^j_{z \bar z}(z_2,w_2)>&=&
[H_0^i+ \frac{<\sum_k h^i_k(z_k,w_1)>}{A}] \times \nonumber\\
&&\times [H_0^j+ \frac{<\sum_{k'} h^j_{k'}(z_{k'},w_2)>}{A} ] 
\eea
In this case factorization and translational invariance follow from
the assumed factorization and translational invariance of $<h^j_k(z,w)>$.
When $w_1=w_2$ the factorization follows from the preceding assumptions
about the v.e.v. of $h^j_k(z,w)$ and from the computation
in Eq.(9)-Eq.(10). \\ 
The occurrence of delta-two can be interpreted as vortices, as I did in
the first section.
The vortex charge is quantized according to
the topological class defined by $\pi_1(SU(N)/Z_N)$.
Alternatively
the gauge fixing $F^{ch}_{z \bar z}$ leaves a residual $U(1)^{N-1}$, and solutions
of the system:
\bea
&& F^{ch}_{z \bar z}(z,w)=0 \nonumber\\
&&i F^0_{z \bar z}(z,w)=H^0_{0} + H^0(z,w;[z_i])
\eea
can be classified by $\pi_1(U(1)^{N-1})$.
If we allow $k$-fold covers of the original surface $Z$,
rational holonomies and vortex charges are allowed at large $N$.
This completes the classification of the contact terms that may occur at large
$N$. In the next section I present a physical interpretation of the structure of
the master field for $F^{0}_{z \bar z}$.

\section{Confinement and the master field}

According to \cite{'t Hooft}, if there is a mass gap,
the phase of pure $SU(N)$-gauge theories
can be classified either as the Higgs or the confining one, depending whether
either the electric or the magnetic fluxes condense, respectively.
If there is no mass gap and the gauge symmetry is unbroken, the theory is in
the Coulomb phase.\\ 
It is quite obvious that if pure $SU(N)$-$YM_4$ were in the Higgs phase,
free vortices could not occur in the master field, since the magnetic charge
is confined in this phase.
Indeed if $F^{0}_{z \bar z}$ is a non-zero constant, the $SU(N)$ gauge group must be
spontaneously broken to the isotropy subgroup of $H^0_0$,
since the gauge orbit under global gauge transformations
is non-trivial.
This case is analogous to the one in which the eigenvalues of a scalar field
in the adjoint representation condense in the vacuum.
This theory looks like a superconductor of type one.\\
If a constant field and a vortex condensate occur at the same time in the
master field for the same eigenvalues, the theory resembles a superconductor
of type two, which can be penetrated by magnetic flux vortices
(they would form lines in 3d and sheets in 4d).\\
One way of seeing this is to look at the equations (for $U(N)$):
\bea
&&F^{\phantom{0}ih}_{\bar z z}=\partial_{[\bar z }
A^{ih}_{z]}+i(A^i-A^h)_{[\bar z }A^{ih}_{z]}+i\sum_{j}A^{ij}_{[\bar z }A^{jh}_{z]}
=0 \nonumber \\
&&i F^{\phantom{0}i}_{\bar z z}=i(\partial_{[\bar z }
A^{i}_{z]}+i\sum_{j}A^{ij}_{[\bar z }A^{ji}_{z]}) 
=-H^{i}
\eea
that appear as a constraint at $N=\infty$ in the functional integral.
Using the ansatz (reduction):
\bea
&& A^{ih}_{z}=A^{ih}_{\bar z }=0~~~~\vert i-h \vert > 1 \nonumber\\        
&& A^{ii+1}_{z}=0 ~~~~\rm{vortex}    \nonumber\\  
&& A^{ii+1}_{\bar z }=0 ~~~~\rm{anti-vortex}~,                    
\eea                                                 
the following Toda equations corresponding to vortices or antivortices are
obtained:
\bea
&& (A^i-A^{i+1})_{z}=-i \partial_{z} \log A^{i i+1}_{\bar z } \nonumber \\
&& -\partial_{z} \partial_{\bar z } \log \vert A^{i i+1}_{\bar z } \vert^2-
2 \vert A^{i i+1}_{\bar z } \vert^2 +  \vert A^{i-1 i}_{\bar z } \vert^2
+ \vert A^{i+1 i+2}_{\bar z } \vert^2 =H^{i+1}- H^i \nonumber \\
&&i \sum_{i} \partial_{[\bar z }A^{i}_{z]}=-\sum_{i}H^i
\eea
and
\bea
&& (A^i-A^{i+1})_{\bar z }=i \partial_{\bar z } \log A^{i i+1}_{z} \nonumber \\
&&  \partial_{z} \partial_{\bar z } \log \vert A^{i i+1}_{z} \vert^2+
2 \vert A^{i i+1}_{z} \vert^2- \vert A^{i-1 i}_{z} \vert^2
- \vert A^{i+1 i+2}_{z} \vert^2 =H^{i+1}-H^i \nonumber \\
&&i \sum_{i}\partial_{[\bar z } A^{i}_{z]}=-\sum_{i}H^i~.
\eea
Toda equations are a $SU(N)$ generalization of the Liouville equation
($SU(2)$) involving only $3N-1$ generators,
the Cartan generators and the immediately off-diagonal charged generators.\\
Liouville and Toda equations are known to possess vortex solutions.
In fact they are the paradigm of vortex equations \cite{Yaffe}.\\ 
To be more precise, a vortex with magnetic charge:
\beq
\frac{2\pi n^{ii+1}}{k}
\eeq
arises wherever the charged field $A^{ii+1}_{\bar z}$  has a zero of the form:
\beq
h(z,\bar z )^{ii+1}(\bar z - \bar z _{1})^{\frac{n^{ii+1}}{k}}~.
\eeq                                                                     
This corresponds to a pole singularity of the vector potential
in the Cartan subalgebra and a $\delta-$like singularity in the
Abelian field strength.
An anti-vortex corresponds to a zero involving the complex conjugate variable
and component of the connection on the $Z$ surface.
The `order parameter' $A^{i,i+1}_{\bar z}$ for vortices of type $i,i+1$
approaches exponentially, with exponent of order $|H^i-H^{i+1}|^{\frac{1}{2}}$,
its asymptotic value
$H^{i+1}-H^{i}$ from the vanishing value in the centre of the vortex.
Hence in 3d the magnetic flux would be squeezed into long flux tubes of
transverse width of the order $|H^i-H^{i+1}|^{-\frac{1}{2}}$.
In the case where vortices and a non-vanishing zero mode occur in the master field,
the v.e.v. of $F^{0}_{z \bar z}$ is in general still different from zero
and the symmetry is broken, unless there is enough magnetic flux to
compensate the constant part and the v.e.v. of  $F^{0}_{z \bar z}$ is zero.
In this last case the symmetry is unbroken, and the superconductor is at its
transition point with the Coulomb phase. \\
There is only one possibility left.
The constant part of the master field vanishes and the magnetic vortices
condense in the vacuum: this is the confining phase of $YM_4$. 
Hence at large $N$, in the gauge $F^{ch}_{z \bar z}=0$, the $SU(N)$
functional integral must be localized on the moduli space of flat connections
with punctures, if the theory confines the electric charge.

\section{Acknowledgements}

I would like to thank Camillo Imbimbo, Giorgio Parisi and Massimo
Testa for several clarifying discussions.

\end{document}